# FPGA-Based Multiplier with a New Approximate Full Adder for Error-Resilient Applications

Ali Ranjbar
*Department of Electrical Engineering, Shiraz Branch Islamic Azad University*
Shiraz, Iran
a.randjbarr@gmail.com

Elham Esmaeili
*Department of Electrical Engineering, Shiraz Branch Islamic Azad University*
Shiraz, Iran
elham.esmaeili421990@gmail.com

Roghayeh Rafieisangari
*Department of Electrical Engineering, Shiraz Branch Islamic Azad University*
Shiraz, Iran
shabnam.rafiei1996@gmail.com

Nabiollah Shiri
*Department of Electrical Engineering, Shiraz Branch Islamic Azad University*
Shiraz, Iran
na.shiri@iau.ac.ir

*Abstract*— Electronic devices primarily aim to offer low power consumption, high speed, and a compact area. The performance of very large-scale integration (VLSI) devices is influenced by arithmetic operations, where multiplication is a crucial operation. Therefore, a high-speed multiplier is essential for developing any signal-processing module. Numerous multipliers have been reviewed in existing literature, and their speed is largely determined by how partial products (PPs) are accumulated. To enhance the speed of multiplication beyond current methods, an approximate adder-based multiplier is introduced. This approach allows for the simultaneous addition of PPs from two consecutive bits using a novel approximate adder. The proposed multiplier is utilized in a mean filter structure and implemented in ISE Design Suite 14.7 using VHDL and synthesized on the Xilinx Spartan3-XC3S400 FPGA board. Compared to the literature, the proposed multiplier achieves power and power-delay product (PDP) improvements of 56.09% and 73.02%, respectively. The validity of the expressed multiplier is demonstrated through the mean filter system. Results show that it achieves power savings of 33.33%. Additionally, the proposed multiplier provides more accurate results than other approximate multipliers by expressing higher values of peak signal-to-noise ratio (PSNR), (30.58%), and structural similarity index metric (SSIM), (22.22%), while power consumption is in a low range.

*Keywords—approximate computing, approximate full adder, multiplier, mean filter.*

## I. Introduction

In real-time application systems, the key objectives are speed, power efficiency, and area optimization. Multiplication and addition are crucial components in digital signal processors (DSPs), central processing units (CPUs), and digital filters [1]. Multipliers serve diverse functions, influenced by the specific constraints of each application. Therefore, it is important to analyze the area, power consumption, and delay characteristics of the multipliers utilized in signal processing tasks [2]. In numerous error-tolerant applications, like multimedia, image processing, and machine learning, exact computations are not always required [1]-[4], and approximate computing is an effective approach. Approximate computing reduces power consumption and improves the performance of embedded systems. By allowing some errors in the outputs of a complex circuit, the logic expressions are simplified, which in turn decreases the logic counts. Approximate-based arithmetic cells require fewer logic gates, resulting in lower power consumption but sacrificing accuracy. Recently, researchers have designed various arithmetic circuits [4], such as full adders (FAs), subtractors, compressors, and multipliers. The FA is a crucial component in arithmetic cells. Utilizing FAs and compressors to add partial products (PPs) simplifies the circuit design. However, replacing exact FA with approximate alternatives can yield significant benefits in circuit performance. In high-speed multipliers, both compressors and FAs facilitate faster accumulation of PPs. In [5], an approximate 4:2 compressor was proposed for adding PPs, offering enhanced error metrics. However, this approach leads to a substantial increase in circuit area. The authors utilized both their proposed exact compressors and approximate compressors to create configurable dual-quality multipliers. However, their approach does not apply to field programmable gate arrays (FPGAs) because current commercial FPGAs lack power gating capabilities. Moreover, their 4:2 compressor still requires significant hardware resources and power, resulting in moderate error rates when implemented on FPGAs. Based on the discussions above, it is evident that using the presented approximation techniques for developing various types of multipliers leads to improvements in performance, energy efficiency, and area reduction. However, controlling their application is challenging. Therefore, approximation techniques that have been successfully applied in application-specific integrated circuits (ASICs) yield limited advantages when transferred to FPGAs. Xilinx and Intel FPGAs offer fast DSP-based multipliers suitable for low-power digital signal processing applications; however, utilizing these multiplier intellectual properties (IPs) leads to substantial routing delays because they are only accessible at certain locations on the FPGA. Recently, [6] proposed optimizing the multipliers by removing the least significant PP of a 4×2 multiplier to conserve look-up tables (LUTs), subsequently using this component to construct larger operand size multipliers. This strategy, however, provided only marginal gains in area and power efficiency. To address these challenges, this research focuses on the design and analysis of a new approximate FA, which is applied in a multiplier and presents an approximate multiplier with various accuracy levels. The introduced approximate multiplier is optimized for effective implementation on FPGAs, achieving high electrical performance characteristics such as low power consumption, minimal area, and low latency, along with minimal accuracy loss. Consequently, this multiplier is well-suited for digital signal-processing applications like image processing. The multiplier is compared with state-of-the-art designs in terms of delay, power consumption, area, and power-delay product (PDP).

This paper is arranged as follows: Section II presents the proposed circuits. Section III analyzes and implements the new approximate designs and error analysis. Section IV discusses application and performance evaluation, and Section V concludes the paper.

## II. PROPOSED CIRCUITS

### A. Presented Approximate FA

FA is a core unit for the performance enhancement of digital systems. Various FAs' techniques are utilized in intermediate modules to generate sum and carry outputs [7]-[8]. A primary limitation of FAs is their operational speed, which necessitates a focus on minimizing delay. Approximate computing offers a trade-off between accuracy and improvements in circuit parameters. Many applications that require significant computational resources are inherently error resilient, given the limitations of human visual perception or the absence of a definitive correct answer for specific problems. Thus, approximate computing can effectively enhance the digital hardware specifications like area, power, and speed for these error-tolerant applications. In this paper, an approximate FA is proposed. As shown in Fig. 1, according to the truth table, the outputs for Sum and $C_{out}$ of the FA exhibit 4 and 2 errors, respectively. Also, the gate-level structure is just one OR gate, and an 8-bit ripple carry adder (RCA) is considered for the performance evaluation. Here, the Sum is generated with only an OR gate and $C_{out}$ equal to B input. The functions of the FA are given by (1)-(2).

$$Sum = A + C_{in} \quad (1)$$

$$C_{out} = B \quad (2)$$

The error distance (ED), error rate (ER), and normalized mean error distance (NMED) for the FA are |-1|, 0.5%, and 0.166, respectively. The ED reflects the difference between the exact and approximate values, specifically focusing on the difference between the $C_{out}$ and Sum (CS) outputs.

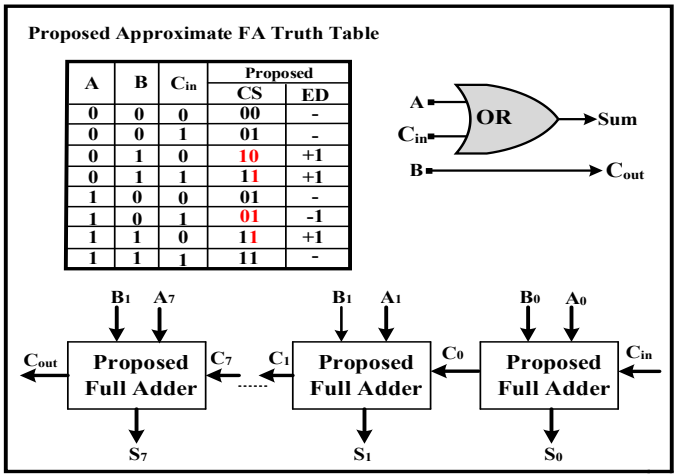

Fig. 1. Truth table, gate level, and 8-bit RCA of the proposed approximate FA.

Therefore, one objective of this FA is to minimize the occurrences of ED=1. Extensive literature indicates that an ED of ⩾ 2 can significantly impair the overall accuracy of a circuit, particularly in more complex structures such as multipliers. But in the proposed circuit ED=1 leads to a lower NMED and indicates better accuracy. Fewer gates through critical paths without inverters, reduce static power and delay. In the 8-bit RCA of Fig.1, the number of approximate bits (NABs) varies to evaluate the performance of the approximate FAs. NAB1 indicates that an approximate FA is applied solely to the least significant bit (LSB). For NAB2, approximate FAs are used for the two LSBs, and this pattern continues up to NAB8, where they are applied to the most significant bit (MSB).

### B. Approximate Multiplier

Multiplier is a fundamental arithmetic operation in various applications like finite impulse response (FIR) filters, discrete cosine transform (DCT), fast Fourier transform (FFT), and multimedia processing. To achieve optimal quality and accuracy in the output data of signal processing modules, it is essential to enhance the speed, area, and power efficiency of the arithmetic modules [9-11]. Consequently, the multiplication module is crucial for reducing computation delay and improving system speed [12]. Fig. 2 illustrates the architecture of a typical 8×8 multiplier, which performs multiplication in three primary steps: (1) recording and generating PPs, (2) reducing PPs, and (3) accumulating those PPs.

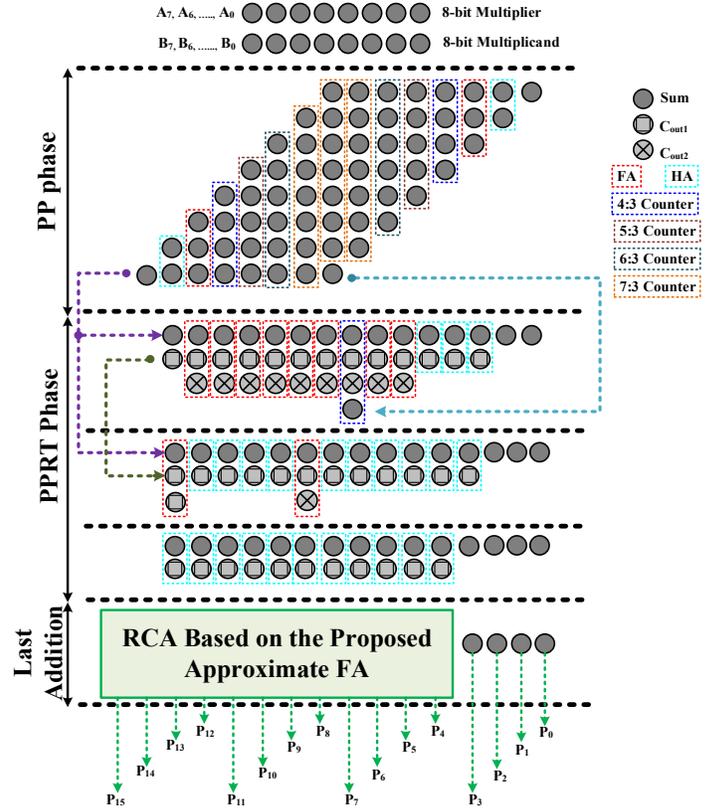

Fig. 2. The architecture of the proposed 8*8 approximate multiplier.

Multipliers can be categorized based on their partial product reduction (PPR) methods, with common types being linear array multipliers and tree multipliers. In an array multiplier, the multiplication of two binary numbers follows an add-and-shift method, which involves a regular structure. However, when dealing with a large number of bits, this type of multiplier can incur significant delays and high power consumption due to carry propagation, highlighting the importance of optimizing carry propagation speed (the critical output path) [13]-[14]. Tree multipliers, organize PPs in rows or columns, reducing the number of components compared to array types. The accumulation of PPs often limits the multiplication speed. An array multiplier employing a modified FA-based multiplexer was developed to minimize power consumption [15]. A multiplier that incorporates an approximate 4:2 compressor demonstrates significantly reduced delay and area compared to other types of multipliers [16], although it does produce a notable error rate. Given these findings, it is critical to focus on the accumulation process to enhance speed, area, and accuracy.

The initial step of multiplication involves generating PPs through logical AND operations. For an 8×8 multiplier with multiplier A (A0 to A7) and multiplicand B (B0 to B7), the first step involves ANDing the LSB of multiplicand B (B0) with every bit of multiplier A. The proposed approximate multiplier comprises three stages, each utilizing different sizes of adders. The first stage contains two half adders (HAs), two proposed approximate FAs, and several counter structures (4:3, 5:3, 6:3, and three 7:3 counters) [12]. The outputs from this stage are passed into the next stage's adders. The second stage includes three HAs and nine FAs, where the outputs from the first stage are added concurrently, with results sent to a final RCA, which represents the final product of the 8×8 multiplication. By replacing the 4:2 compressor in the proposed multiplier with approximate FAs, fewer gates than traditional structures are required. This reduction contributes to lower power consumption and enhanced speed, area, and accuracy. So, in the proposed multiplier, the use of approximate FAs instead of 4:2 compressors is preferred for minimizing critical path delays. By strategically selecting the appropriate combination of counters and FAs, the critical paths can be further optimized, leading to improved overall performance.

### III. SIMULATION RESULTS

The proposed approximate circuits and references are described using VHDL and are synthesized and implemented using the Xilinx ISE Design Suite 14.7 on the Xilinx Spartan3 XC3S400-4PQ208 FPGA board to verify its functionality. For delay and power calculations and simulations, ISE XPower Analyzer and ISim simulator are used. The results are acquired under standard operating conditions with a system clock frequency of 50 MHz. Each adder and multiplier are simulated and verified separately. The circuitry performance is evaluated by PDP, and the accuracy is checked by NMED. The results of normalized PDP and NMED of the introduced FA and references for the NAB1 in the RCA are shown in Fig. 3. Regarding PDP, the proposed FA has the minimum value and shows the best performance in terms of power and NMED. Let's examine an N×N multiplier. To assess the quality of the multiplier, the error metrics of mean error distance (MED), mean relative error distance (MRED), and NMED are used. $ED_i$ represents the ED, which is the arithmetic difference between the i-th accurate product and its approximate counterpart, while $S_i$ denotes the i-th accurate product. The definitions of MED, MRED, and NMED are as follows:

$$MED = \frac{1}{2^{2N}} \sum_{i=1}^{2^{2N}} ED_i \qquad (3)$$

$$MRED = \frac{1}{2^{2N}} \sum_{i=1}^{2^{2N}} \frac{ED_i}{S_i} \qquad (4)$$

$$NMED = \frac{1}{(2^N-1)^2} \times \frac{\sum_{i=1}^{2^{2N}} ED_i}{2^{2N}} \qquad (5)$$

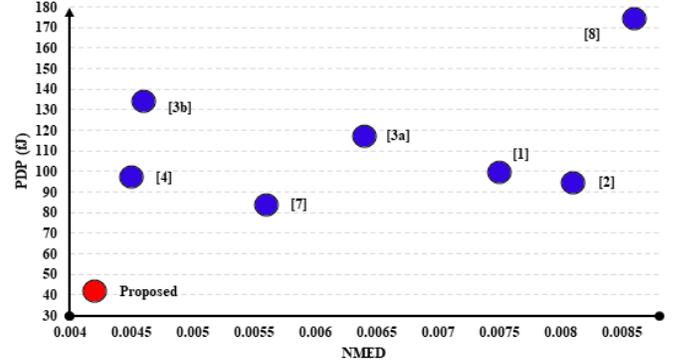

Fig. 3. Results of the PDP and NMED for proposed FA and references.

Table. I show the comparison between the presented 8×8 multiplier and previous works in terms of resource utilization and accuracy. The proposed 8×8 multiplier outperforms the prior works utilizing 71 out of 7168 available 4-input LUTs (0.99%) and 9 out of 7168 flip-flops (0.13%). Compared to [6], the proposed multiplier can achieve power and PDP improvements of 56.09% and 73.02%, respectively. As shown in Table. I, in terms of MED, MRED, and NMED proposed multiplier has better performance and less resource utilization compared to other references.

TABLE. I. PERFORMANCE ANALYSIS OF THE 8 × 8 MULTIPLIERS.

| Design | Power (mW) | Delay (ns) | No. of LUTs | PDP (PJ) | MED | MRED | NMED |
|---|---|---|---|---|---|---|---|
| M [13] | 0.636 | 5.31 | 68 | 3.377 | 0.0034 | 0.3676 | 0.0154 |
| M [14] | 0.468 | **4.84** | 71 | 2.265 | 0.0022 | 0.0196 | 0.0034 |
| M [15] | 0.744 | 5.11 | 101 | 3.802 | 0.0011 | 0.0548 | 0.0154 |
| M [6] | 0.984 | 8.31 | **57** | 8.177 | 0.0054 | 0.0029 | 0.0008 |
| M [16] | 0.780 | 5.37 | 91 | 4.189 | 0.0013 | 0.0062 | 0.0020 |
| **Proposed** | **0.432** | 5.11 | 71 | **2.207** | **0.0010** | **0.0148** | **0.0017** |

### IV. APPLICATION

One of the best possible ways for multiplier circuits' assessments is their applications in image processing. In [17], an image enhancement algorithm was implemented on FPGA, which highlights the potential of FPGA-based systems in image processing applications, especially denoising images. Noise is created by interferences that may occur during the image acquisition and transmission stages in a digital image. A noisy image can be modeled as follows:

$$g(x,y) = f(x,y) + \eta(x,y) \qquad (6)$$

The noisy image, g(x, y), consists of the original image (f (x, y)) and the performed noise on it (η(x, y)). There are many different models for image noise like Gaussian, Rayleigh, Erlang, Exponential, Uniform, Bumpy, and Salt and pepper noise. In this paper, the Gaussian noise is applied to the input images and then, a system including the mean filter algorithm reduces the noise of the images and gives the output image with the least possible noise. The mean filter is one of the spatial filters that is utilized in smoothing, denoising, and restoration of images in digital image processing. The mean filter can be used to remove various types of noise and is calculated as follows:

$$J(x,y) = \sum_{i=-k}^{k}\sum_{i=-k}^{k}\frac{1}{(2k+1)^2}I(x+i,y+j) \quad (7)$$

The mean filter first considers a window around a pixel and then takes the average intensity of the pixels in that window as the new value of that pixel. Usually, the window around a pixel is considered to be a square that has $(2k+1)$ pixels on each side of itself. If it is the original image the intensity of $(x,y)$ pixel of this image is $I(x,y)$, then a mean filter with a $(2k+1)\times(2k+1)$ window, changes the intensity of $(x,y)$ pixel from $I(x,y)$ to $J(x,y)$. As can be seen, the mean filter is a linear filter with a $(2k+1)\times(2k+1)$ matrix mask that all arrays of the mask are $w_{i,j} = \frac{1}{(2k+1)^2}$. In this paper, the image processing algorithm is implemented based on a 3×3 window of the input image pixels and a 3×3 window of the mask of the mean filter. As shown in Fig. 4, each pixel of the mean filter mask is $\frac{1}{9}$. In the pre-processing stage, each image is affected by the Gaussian noise with 0.003 variances. Then the grayscale images are converted to binary images as the single-bit inputs of the system.

The proposed design is performed on the Xilinx Spartan3 XC3S400-4PQ208 FPGA board. To implement the mean filter utilizing the proposed multiplier, each 3×3 window of the pixels of the input image should be multiplied by the mean filter mask using the proposed 8×8 multiplier and then, added together to reach the mean value of the centered pixel of the input image. Fig. 4 illustrates how the proposed multiplier is utilized in the mean filter algorithm to reduce the noise of images.

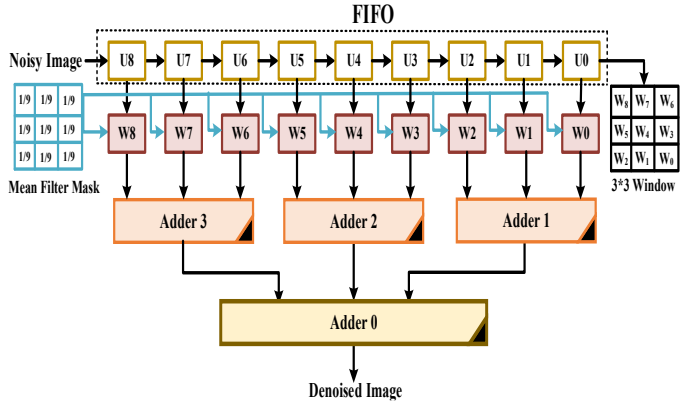

Fig. 4. Hardware implementation of the mean filter algorithm using the proposed multiplier on an FPGA.

All output pixels of the system are sorted respectively to make the denoised image out of a noisy image. As mentioned before, to multiply each pixel and eight pixels surrounding it at the mean filter mask, the pixels of the input image should be in 3×3 format and a window. To do so, a first-in-first-out (FIFO) memory with nine flip-flops is used in this design, as shown by U8, U7, …, and U0 in Fig. 4. At first, each pixel of the image is entered the first FIFO block (F8), stays as long as a clock in the flip-flop, and then transfers to the next FIFO block (F7) until it quits from the last block (F0). In every rising edge of the clock, the binary pixel in each FIFO block is copied eight times as it can be one of the inputs of the proposed 8×8 multiplier. On the other hand, the other input of each multiplier is $\frac{1}{9}$ as a pixel of the mean filter mask. With nine FIFO blocks, nine multipliers are needed to apply the mask to the pixels that are shown as W8, W7,…, and W0 in Fig. 4. Eventually, the outputs of the multipliers are added to each other using four adders of the FPGA (Adder 3,…, and Adder 0). The usage of four adders instead of just one is the limited input/output block of the FPGA device. However, it is obvious that the order of the pixels is not affected and the outputs of the system show the denoised image correctly. The best output of the system shows that the implemented mean filter algorithm by using the proposed approximate multiplier reduces the noises from the input images and smooths the images by an acceptable range. As is shown in Table. II, both sensitive (bioimages) and non-sensitive (standard) images are denoised and smoothed with less power, and the best peak signal-to-noise ratio (PSNR) and structural similarity index metric (SSIM).

TABLE. II. THE PROPOSED IMAGE PROCESSING RESULTS.

| Gray Inputs | Noisy Inputs | Denoised Output | PSNR (dB) | SSIM | Power (mW) | Power Saving (%) |
|---|---|---|---|---|---|---|
| | | | 49.12 | 0.95 | 1.23 | 33.32 |
| | | | 49.32 | 0.98 | 1.21 | 33.33 |
| | | | 39.99 | 0.88 | 1.25 | 33.31 |
| | | | 39.52 | 0.88 | 1.27 | 33.31 |
| | | | 39.45 | 0.87 | 1.29 | 33.30 |

By substituting the exact multipliers with the designed approximate multipliers, the mean filter performance is assessed. The quality metrics including PSNR, SSIM, power, and power saving are evaluated. As provided in Table. III, the proposed approximate multiplier is a highly precise design with power savings of 33.33%.

TABLE. III. QUALITY AND POWER SAVINGS COMPARISON FOR 8×8 MULTIPLIERS.

| Multiplier | PSNR (dB) | SSIM | Power (mW) | Power Savings (%) |
|---|---|---|---|---|
| [13] | 15.98 | 0.77 | 1.64 | 14.2 |
| [14] | 44.98 | 0.99 | 1.38 | 4.2 |
| [15] | 35.14 | 0.99 | 1.87 | 30 |
| [6] | 14.63 | 0.81 | 1.92 | 33.30 |
| [16] | 50.51 | 0.99 | 2.13 | 28.3 |
| **Proposed** | **50.62** | **0.99** | **1.20** | **33.33** |

Fig. 5, demonstrates SSIM and PSNR for several approximate multipliers, where the proposed multiplier is more accurate than the others. Note that concerning the SSIM, the proposed designs, and [16], are the best (highest SSIM values). Notably, the PSNR is calculated by (8).

$$PSNR = 10\, log_{10} \frac{2552^2}{MSE} \qquad (8)$$

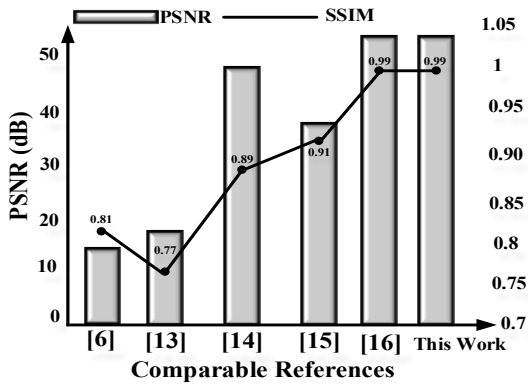

Fig. 5. PSNR and SSIM for the image processing.

The results of the FPGA implementation of the multiplier and mean filter system confirm the research contribution in image processing, especially sensitive images like medical magnetic resonance imaging (MRI), and computed tomography (CT) scans. On the other hand, the presented system is a useful method for disease detection and low-power neural network implementation.

## V. CONCLUSION

An approximate full adder (FA)-based 8×8 multiplier is presented and implemented on the field programmable gate array (FPGA). The multiplier uses a new approximate FA to reduce hardware complexity, delay, and power. The multiplier outperforms look-up table (LUT)-based multipliers available on FPGAs regarding dynamic power dissipation, power-delay-product (PDP), and mean relative error distance (MRED). The results implementing the multiplier in the mean filter system show that it achieves power savings of 33.33%. Additionally, the proposed multiplier produces more accurate output than other approximate multipliers by achieving higher quality in terms of peak signal-to-noise ratio (PSNR) and structural similarity index metric (SSIM) while consuming less power.

## AI USAGE STATEMENT

The authors confirm that no artificial intelligence tools were used in the preparation of this article.